\documentstyle[12pt]{article} 
\begin{document}

\title{{\bf Field equations or conservation laws?}}
\author{{\large  Mauro Francaviglia, Marcella Palese and Ekkehart Winterroth}
\\ 
{\small Department of Mathematics, University of Torino, Italy
}\\ 
{\small Via C. Alberto 10 -- 10123, Torino}}
\date{} 
\maketitle 

\begin{abstract}

We explicate some epistemological implications of stationary principles and in particular of Noether Theorems. Noether's contribution to the problem of covariance, in fact, is epistemologically relevant, since it moves the attention from equations to conservation laws.

\medskip 


\end{abstract}

\section{Field equations dual to stationary principles} \indent 

We are interested in the study of the relation between symmetries (i.e., invariance properties) of field equations and corresponding conservation laws, more precisely, in the investigation of some aspects concerning the interplay between symmetries, conservation laws and variational principles. 

As is well known, the theory of General Relativity appeared after Albert Einstein's [$1879-1955$] struggle, during the years $1912-1914$\footnote{Within the collaboration with Marcel Grossmann [1878-1936].}, whether founding his theory on the covariance of field equations or rather on the covariance of the conserved quantities (Einstein and Grossmann $1913$). David Hilbert [$1862-1943$] also dedicated his study to this important question as testified by Emmy (Amalie) Noether [$1882-1935$] in the end of her famous and celebrated paper Invariante Variationsprobleme \cite{(Noether 1918)} in the section titled `{\em Eine Hilbertsche Behauptung}'\footnote{Noether (1918) p. 253}.

In this short note we intend to underline some overlooked, however fundamental, aspects and implications of E. Noether's contribution to the problem of energy and in general of conserved quantities in field theories. Noether's contribution to the problem of covariance, in fact, is epistemologically relevant, since it moves the attention from equations to conservation laws, {\em founding the theory on the invariance of the action} (i.e., of the Lagrangian).

Accordingly, in a very recent paper \cite{(Francaviglia et al. 2013)} we studied Noether conservation laws associated with  some `variational' invariance of global Euler-Lagrange morphisms associated with local variational problems of a given type. In this context, the question arises whether we should be interested in conservation laws different from those directly associated with invariance properties of field equations. The answer to this question relays on Emmy Noether's paper.

As is well known, in fact it was motivated by the fact that, although the gravitational field equations were global, the associated conservation laws found by Einstein by a nonvariational approach were not (think of the well known energy-momentum {\em pseudo-tensor}). Explicitly, in the introduction of her paper, Noether  wrote:

{\em \"Uber diese aus Variationsproblemen entspringenden Differentialgleichungen lassen sich viel pr\"azisere Aussagen machen als Ÿber beliebige, eine Gruppe gestattende Differentialgleichungen, die den Gegenstand der Lieschen Untersuchungen bilden.}\footnote{`Concerning these differential equations that arise from problems of variation, far more precise statements can be made than about arbitrary differential equations admitting of a group, which are the subject of LieÕs researches'. Translated from German by M.A. Tavel in \cite{(Tavel 1971)}.}

The relevance of the study of differential equations generated by an invariant  variational problem in its whole is in the issue of a major refinement in the results: to symmetries of equations could correspond conservation laws which have a nonvariational meaning and thus could not be characterized in a similar  precise manner.

Sometime it is improperly stated that Noether's Theorem  would be `formulated for Euler-Lagrange equations in field theory'. Instead, it is important to stress that Noether's I and II Theorems actually are statements about the invariance of a variational problem with respect to a finite continuous group of transformations and an infinite continuous group of transformations, respectively. The direct object of Noether's investigations are what she calls `{\em Lagrangeschen Ausdr\"ucke; d.h. die linken Seiten der Lagrangeschen Gleichungen}' \cite{(Noether 1918)}, which we shall call hereafter Euler-Lagrange expressions. {\em The accent is not put on field equations} although her results have, of course, also consequences concerning invariance properties of equations. It is maybe noteworthy that all Noether considerations are made off shell, i.e., not along solutions of Euler-Lagrange equations. It is also important to stress that Noether immediately considers the formulation of a variational problem at the infinitesimal level  of `{\em integralfreie Identit\"at}'.

Noether's Theorem II is in fact concerned with a variation of the Euler-Lagrange expressions. Symmetry properties of the Euler-Lagrange expressions play a fundamental role since they  introduce a cohomology class which adds up to Noether currents\footnote{A formulation in modern language of Noether's results.}; they are related with invariance properties of the first variation, thus with the vanishing of a second variational derivative. The concept of a variation of Noether current is then clearly involved. In line with Lepage's cornerstone papers \cite{(Lepage 1936)}, which pointed out the fact that the Euler-Lagrange operator is a quotient morphism of the exterior differential, we consider a geometric formulation of the calculus of variations on fibered manifolds for which the Euler-Lagrange operator is a morphism of a finite order exact sequence of sheaves. The module in degree $(n+1)$, contains so-called  (variational) dynamical forms; a given equation is globally an Euler-Lagrange equation if its dynamical form is closed in the complex of global sections (Helmholtz conditions) and its cohomolgy class is trivial.

Therefore, symmetries of Euler-Lagrange morphisms or, more generally, of so-called variational `dynamical forms' are considered insomuch as they can provide informations about Noether currents of some potential Lagrangians, also in the spirit of the Bessel-Hagen version of Noether's Theorem II \cite{(Bessel-Hagen 1921)}. 

\subsection{{\em Being} and {\em Becoming}\,: a contemporary perspective}

As we said, Emmy Noether clearly pointed out how, considering invariance of variational problems, a major refinement in the description of associated conserved quantities is achieved.

In spite of the enormous amount of literature on the applications of the Noether Theorem I (Markov processes, engineering, material sciences, signal propagation and so on), lesser study instead is dedicated to the second part of Noether Theorem. Natural and gauge-natural classical description of field theory is given\footnote{Without quoting the original strictly related Noether's result!} in terms of generalized Bianchi identities by physicists (see e.g. Peter G. Bergmann [1915-2002]), thus revealing an underlying epistemological position aiming to give relevance to equations (although variational) rather than to conservation laws \cite{(Bergmann 1949)}. The ensuing problem of the non covariance of the latter seems to have (according with Einstein final epistemological position) a secondary importance: conserved quantities can be always somehow suitably `covariantized' (think of the Komar superpotential), e.g., introducing some kind of background. Therefore, the epistemological relevance of Noether's approach to the study of field theory has not yet been completely uncovered in all its implications. 

In order to explicate them, in \cite{(Francaviglia et al. 2012)} considering Noether conservation laws associated with the invariance of global Euler-Lagrange morphisms generated by local variational problems of a given type, we introduced the {\em new concept of conserved current variationally associated with locally variational invariant field equations}. 

As already stressed, in order {\em to understand the structure of a phenomenon described by field equations}, one should be interested in conservation laws more precisely characterized than those directly associated with invariance properties of field equations. Thus, it is of fundamental importance to seek for conservation laws coming from invariance properties of a (possibly local) variational problem in its whole (rather than a field equation solely) to find a way of associating global conservation laws with a local Lagrangian {\em field} theory generating global Euler-Lagrange equations.

From Physics' point of view, {\em field equations} appear to be a fundamental object, since they {\em describe the changing} of the field in base space. Somehow, physicists are generally well disposed to give importance to symmetries of equations, because they are transformations of the space leaving invariant the description of such a change which is provided by means of field equations. 

On the other hand the possibility of formulating a variational principle (i.e., a principle of {\em stationary} action) - from which both changing of fields and associated conservation laws (i.e., quantities not changing in the base space) could be obtained  - has been one of the most important achievements in the history of mathematical and physical sciences in Modern Age.  It allows, in fact, to keep account of {\em both} what ({\em and how}) changes and what ({\em and how}) is conserved. In the variational calculus perspective, we could say that  Euler-Lagrange field equations are `adjoint' to stationary principles up to conservation laws: a contemporary mathematical formulation of the duality between {\em Being} and {\em Becoming}.

\section{Conclusion}

In a famous paper sir M. Atiyah \cite{(Atiyah 2005)} wrote: 

{\em 
Standard text-books make great play with the technical details, introducing coordinates, writing equations and then showing that the resulting physics is independent of the choice of coordinates. To a geometer this is perverse. The fundamental link is from physics to geometry, from force to curvature and the algebraic machinery that encodes this is secondary. God created the universe without writing down equations!
}

Thus if `god created the universe without writing down equations' the epistemological position of choosing invariants versus equations can be even strengthened: we assume the field to be described by a strong stationarity condition, requiring more than the invariance of the action,  even the invariance of the first variation: this enables us to define a global conserved current associated with invariant field equations. 

In this perspective in \cite{(Francaviglia et al. 2013)} we found that the conserved current associated with a generalized symmetry, assumed to be also a symmetry of the variational derivative of the corresponding local inverse problem, is variationally equivalent to the variation of the strong Noether currents for the corresponding local system of Lagrangians. Moreover, if the variational Lie derivative of the local system of Lagrangians is a global object, such a variation is variationally equivalent to a global conserved current.

\end{document}